\documentclass[journal=iecred, manuscript=article]{achemso}
\usepackage[T1]{fontenc} 
\usepackage{graphicx,epsf, fancyhdr,tree-dvips,xcolor}  
\usepackage{natbib, cprotect}
\usepackage{amssymb, psfrag, units, booktabs, algorithm}
\usepackage{amsmath}
\usepackage{placeins}
\usepackage{eurosym, color, xfrac}
\def\ve#1{\mathchoice{\mbox{\boldmath$\displaystyle#1$}}
	{\mbox{\boldmath$\textstyle#1$}}
	{\mbox{\boldmath$\scriptstyle#1$}}
	{\mbox{\boldmath$\scriptscriptstyle#1$}}} 
\newcommand{\dd}{\text{d}}
\newcommand{\dt}{\dd t}
\newcommand{\ff}{\text f}

\newcommand{\pp}{\text p}
\newcommand{\eg}{\text{e.g., }}
\newcommand{\ie}{\text{i.e., }}
\newcommand{\wrt}{\text{w.r.t. }}

\newcommand{\st}{\text{s.t.}}

\usepackage{multirow}
\newcommand{\treetextsize}[1]{{\footnotesize #1}}
\mciteErrorOnUnknownfalse

\author{Radoslav Paulen}
\email{radoslav.paulen@stuba.sk}
\phone{+421 (0)259 325 730}
\fax{+421 (0)259 325 340}
\author{Miroslav Fikar}
\affiliation[STUBA]{Faculty of Chemical and Food Technology, Slovak University of Technology in Bratislava, Bratislava, Slovakia}

\title{Dual-control based approach to batch process operation under uncertainty based on optimality-conditions parameterization}


\begin{document}
\begin{abstract}
This paper presents a scheme for dual robust control of batch processes under parametric uncertainty. The dual-control paradigm arises in the context of adaptive control. A trade-off should be decided between the control actions that (robustly) optimize the plant performance and between those that excite the plant such that unknown plant model parameters can be learned precisely enough to increase the robust performance of the plant. Some recently proposed approaches can be used to tackle this problem, however, this will be done at the price of conservativeness or significant computational burden. In order to increase computational efficiency, we propose a scheme that uses parameterized conditions of optimality in the adaptive predictive-control fashion. The dual features of the controller are incorporated through scenario-based (multi-stage) approach, which allows for modeling of the adaptive robust decision problem and for projecting this decision into predictions of the controller. The proposed approach is illustrated on a case study from batch membrane filtration.
\end{abstract}

\section{Introduction}
\label{sec:introduction}
Optimization technology has deeply penetrated into the state-of-the-art approaches to process design and operations. This is also true for the class of batch processes, whose optimal operation is a rich field of research. After the early efforts~\citep{boc84, cut89, vas94a} devoted to solution of the transient (dynamic) optimization, several effective software solutions exist today (e.g.,\cite{bar93, gproms, dynopt, Houska2011a, andersson2012}).

The optimization-based solutions are inevitably based on mathematical models. At present, a typical challenge is to handle uncertainty present in the models. In the batch process operation, one of the main goals is to reduce the variability among the produced batches despite the uncertainties. This problem has struck the attention of many research groups~\citep{nag03, sri03b, ade09, stu11, fra13, luc13, mar15, jan16, hou17}.

In this paper, we consider a real-time implementation of a controller that solves the dynamic optimization problem of the form:
\begin{subequations}\label{eq:prob_gen}
\begin{align}
\min_{u(t)\in [u^L, u^U], t_\ff} & \ \mathcal J(\ve p) \!:= 
\min_{u(t)\in [u^L, u^U], t_\ff} \int_{0}^{t_\ff} \!F_0(\ve x(t, \ve p), \ve p) + F_u(\ve x(t, \ve p), \ve p) u(t)\,\dt\\
\st \ & \dot{\ve x}(t, \ve p) = \ve f_0(\ve x(t, \ve p), \ve p)
+ \ve f_u(\ve x(t, \ve p), \ve p) u(t),\label{eq:model}\\
& \ve x(0) = \ve x_0, \quad \ve x(t_\ff, \ve p) = \ve x_\ff, 
\end{align}
\end{subequations}
where $t$ is time with $t\in[0, t_\ff]$, $\ve x(t)$ is an $n$-dimensional vector of state variables, $\ve p$ is an $m$-dimensional vector of time-invariant model parameters, $u(t)$ is a (scalar) manipulated variable, $F_0(\cdot)$, $F_u(\cdot)$, $\ve f_0(\cdot)$, and $\ve f_u(\cdot)$ are continuously differentiable functions, $\ve x_0$ represents a vector of initial conditions, and $\ve x_\ff$ are specified final conditions.

We note here that an inclusion of multi-input and/or state-constrained cases is a straightforward extension but it is not considered in this study for the sake of simplicity of the presentation. We also note that the specific class of input-affine systems is a suitable representation for a large variety of the controlled systems~\citep{han06}. For a general nonlinear model, one may use simple tricks to rearrange the model into input-affine structure~\cite{son98}, which might though increase the number states of the problem. In the domain of chemical engineering, it is, however, very common to encounter input-affine problems~\cite{amr10} (e.g., when the optimized variables is reactor feed) or to reformulate the model and arrive at the input-affine structure~\cite{lio95}.

We will assume that a structurally correct mathematical model of the plant is available that describes the plant behavior and that all potential disturbances are precisely measured. We will also assume that all the state variables are measured or that there is an ideal state estimator employed, which converges to the true plant states within one sampling period. In reality, one would use a state estimator whose uncertainty about the state estimates would need to be taken into account. Under these assumptions, the only source of uncertainty is present in the unknown values of model parameters, where a prior knowledge is assumed about the parameters, i.e., the true values of the parameters lie in the a priori known interval box $\ve P_0:=[\ve p_0^L, \ve p_0^U]$, where superscripts $L$ and $U$ denote the lower and the upper bounds of $\ve p$.

The presented problem was studied in several previous works using on-line or batch-to-batch adaptation of the optimality conditions~\citep{fra13} or by design of robust controller for tracking the conditions of
optimality~\citep{nag03}. Recently, several advanced robust strategies were presented in the framework of model predictive control~\citep{luc13, hou17}. This paper proposes adaptation of the aforementioned approaches to the problem of robust optimal control of batch processes.

We base our approach on the parameterization of the optimal controller using the conditions of optimality given by Pontryagin's minimum principle. This step reduces computational burden when projecting the parametric uncertainty in controller performance and feasibility and when solving the problem~\eqref{eq:prob_gen} in the shrinking-horizon fashion in real time. In order to improve the control performance, we use on-line parameter estimation. Finally, we derive the dual controller that considers adaptation of the optimal control inputs based on projected uncertainty and on prediction of the future learning of the controller.

The term dual control was first coined by Feldbaum~\cite{fel60} and subsequently used in the literature (see the surveys~\cite{fil00, unb00}) as a control scheme, where the controller performs optimal decision \wrt probing (excitation) control actions to learn the system behavior and (cautious) actions that drive the still partially uncertain system into the desired operating regime. One main distinguishing feature among the dual control approaches is whether a) the controller explicitly involves the injection of the excitation signals in its design criterion (e.g., a commonly used term in the objective function of optimization-based controllers that weighs the importance of plant excitation and performance) or whether b) the performance-optimal excitation results from the awareness of the controller of the effect of the probing actions on the control performance. The first class of approaches is referred to as explicit~\cite{mil82, han15, hei15, la17, lor19} and the latter one as implicit~\cite{lee09, tan14, tan19, hei17, fen18}. In this work, we will use an implicit approach that is based on multi-stage NMPC framework of Thangavel et al.~\cite{tha15, tha18} since, despite being computationally more demanding as an explicit dual-control strategy, it requires no a priori tuning of the objective regarding the importance of the probing and optimizing control actions. We adapt this method into the shrinking-horizon-based control for batch processes.

The outline of the paper is as follows. In the next section, we introduce the preliminary theoretical knowledge on Pontryagin's minimum principle~\citep{pon62} and on set-membership estimation~\citep{schw68, fog82}. Next we propose two strategies for the implementation of the control policy based on the principles of adaptive and dual control, respectively. Finally, we present a simulation case study from chemical engineering domain and discuss various aspects of the obtained results.

\section{Nominal optimal control}
Using Pontryagin's minimum principle (see Appendix), the optimal control trajectory for the problem~\eqref{eq:prob_gen} is given as a step-wise strategy~\citep{pau15jpc}
\begin{align}\label{eq:singular_ctrl_switch}
u^\ast(t, \ve\pi) :=
\begin{cases}
  u^L, &  t\in[0, t_1], \ S(\ve x(t, \ve p), \ve p) >0,\\
  u^U, & t\in[0, t_1], \ S(\ve x(t, \ve p), \ve p) <0,\\
  u_\text s(\ve x(t), \ve p), & t\in[t_1, t_2], \ S(\ve x(t, \ve p), \ve p) =0,\\
  u^L, & t\in[t_2,t_\ff], \ S(\ve x_\ff, \ve p) <0,\\
  u^U, & t\in[t_2,t_\ff], \ S(\ve x_\ff, \ve p) >0,
\end{cases}\\
\ve x_\ff=\ve x(t_2, \ve p) + \int_{t_2}^{t_\ff}\!\ve f_0(\ve x(t, \ve p), \ve p) + \ve f_u(\ve x(t, \ve p), \ve p) u^\ast(t)\,\dt,
\end{align}
where $\ve\pi:=(\ve p^T, t_1, t_2, t_\ff)^T$ is the vector that parameterizes the optimal control strategy, $S(\ve x, \ve p)$ is the switching function identified by the minimum principle, and the singular control $u_\text s(\ve x(t), \ve p)$ is derived from the switching function (see Appendix). Note that the presented optimal strategy determines implicitly the switching times $t_1$ from saturated to singular control and $t_2$ from singular to saturated control terminal time as well as the terminal time $t_\ff$.

In case that the use of the minimum principle turn out to be too complex (e.g., many derivatives are needed to characterize the solution), a recently presented parsimonious input parameterization~\citep{erd18, rod19} can be used.

\section{Real-time implementation of the control scheme}
As the optimal control structure is a function of uncertain parameters of the
process model, the uncertainty should be taken into account when devising a
real-time implementation of the optimal control on the process. We will assume
that the uncertainty is a priori bounded as $\ve p \in \ve P_0:=[\ve p_0^L, \ve p_0^U]$ and has a nominal realization $\ve p_0^0$, which is taken as a mid-point of the interval vector $\ve P_0$.

We will also assume that the parameters can be inferred from the measurements given by a measurement function $\ve y_k=g(\ve x(t_k), \ve p)$. The measurements are assumed to be corrupted by unknown-but-bounded measurement noise. This condition implies the use of set-membership estimation (see Appendix).

We are interested in the determination of parametric bounds such that
\begin{equation}
 \ve P_k \subseteq \ve P_{k-1} \subseteq \dots \subseteq \ve P_1 \subseteq \ve P_0.
\end{equation}
The parametric bounds can be determined through solution of a series of optimization problems as:
\begin{subequations}\label{eq:nlp_par_bounds}
\begin{align}
p_{k,j}^L/&p_{k,j}^U := \arg\min_{\ve p\in\ve P_{k-1}} \ /\arg\max_{\ve p\in\ve P_{k-1}} p_j\\
\text{s.t. }& \dot{\ve x}(t,\ve p) = \ve f (\ve x(t,\ve p), \ve p),\ \forall t\in[t_0,t_k],\\
&\ve x(t_0, \ve p) = \ve h (\ve p),\\
&\hat{\ve y}(t_i, \ve p) = \ve g (\ve x(t_i, \ve p), \ve p),\ \forall i \in \{1, \dots, k\},\\
&-\ve\sigma \leq \hat{\ve y}(t_i, \ve p) - \ve y_i \leq \ve\sigma,\ \forall i \in \{1, \dots, k\}, \label{eq:nlp_par_bounds_err}
\end{align}
\end{subequations}
where $j\in\{1,\dots, n_p\}$ indicates the $j^\text{th}$ element of a vector.

\subsection{Implementation via adaptive robust control}

\begin{figure}
\centering
\psfrag{alpha}{\hspace{-0.4cm}$u^*(t,\cdot)$}
\psfrag{time}{$t$}
\psfrag{0}{\hspace{-0.4cm} $u^L$}
\psfrag{1}{\hspace{-0.5cm} $u^U$}
\psfrag{t1}[t][b][1][0]{\hspace{-0.1cm}$[t^L_1(\ve P_0), t^U_1(\ve P_0)]$}
\psfrag{t2}[t][b][1][0]{\hspace{-0.4cm}$[t^L_2(\ve P_0), t^U_2(\ve P_0)]$}
\psfrag{t3}[t][b][1][0]{\hspace{0.4cm}$[t^L_\ff(\ve P_0), t^U_\ff(\ve P_0)]$}
\includegraphics[width=0.8\linewidth]{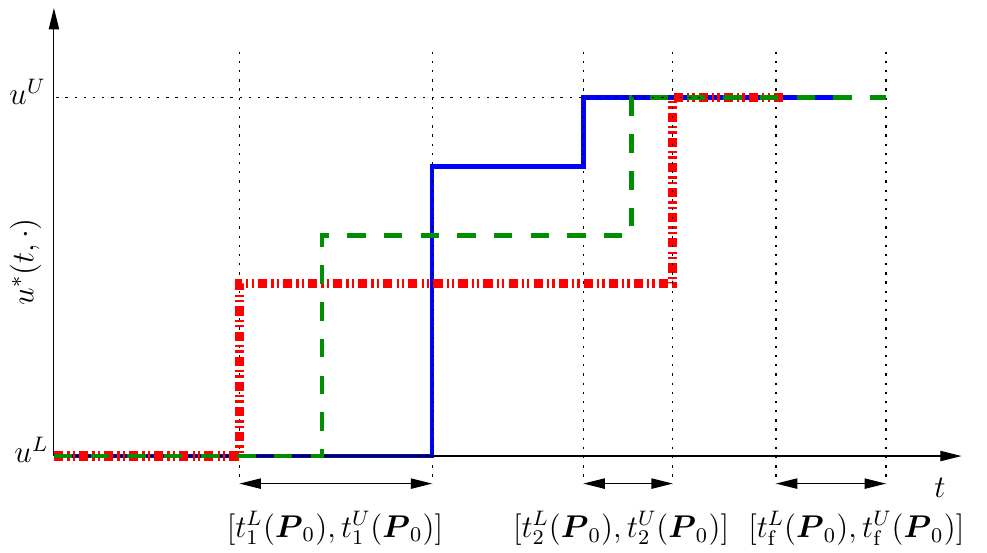}
\vspace{5pt}
\caption{Illustration of the parameterization of the optimal control policy under uncertainty.}
\label{fig:uparam}
\end{figure}

Given the optimal control sequence~\eqref{eq:singular_ctrl_switch}, it is
possible to enclose all the reachable states
$\ve X(t, \ve P) \ni \ve x(t, \ve p),\,\forall \ve p\in\ve P_0$ 
of~\eqref{eq:model}~\citep{vil15} such that one can identify the realization of the optimal control-input sequence (\eg
$u^\ast=\{u^L, u_\text s(\ve x(t, \ve p), \ve p), u^U\}$) and the switching times of the control structure as functions of uncertain parameters $t_1(\ve p)$, $t_2(\ve p)$, and $t_\ff(\ve p), \forall \ve{p}\in \ve P_0$. This is illustrated in Fig.~\ref{fig:uparam} for a simple case, where the singular control is given by a constant varying with parameters. The identified lower and upper bounds are shown for the switching times and for the value of singular control. Later in the text, we will adopt a short-hand notation $[t_i], \ \forall t_i\in\{t_1, t_2, t_\ff\}$ for the uncertain switching times.

At this point, one can formulate a semi-infinite program similar to~\cite{stu11} or some related problem (\eg using polynomial expansion~\citep{hou17}), to determine the parameters of the optimal control structure that lead to the best performance in the worst case. This, however, might lead to an overly conservative strategy. In order to reduce the conservatism, parameter estimation can be used for exploitation of data gathered along the process run~\citep{ade09, lucia2014b}. In case the set-membership estimation is employed, the knowledge about the uncertain parameters can be updated in each sampling instant of the plant, $k>0$. One can then solve
\begin{subequations}\label{eq:prob_rob_nco}
\begin{align}
 \min_{t_1, t_2, t_\ff} & \ \|\mathcal J(\ve P_k) - \mathcal J(\ve p^0_k)\|_2^2\\
 \st \ & \dot{\ve x}(t) = \ve f_0(\ve x(t), \ve p)
         + \ve f_u(\ve x(t), \ve p) u^\ast(t, \ve\pi),\ \forall \ve p \in \ve P_k,\ \forall t\in[t_k, t_f]\\
 & \ve x(t_\ff, \ve P) \ni \ve x_\ff,
\end{align}
\end{subequations}
for some given initial conditions $\ve x(t_k, \ve p) = \ve x_k$ and $\ve P_k$, where we propose to minimize the variance of the objective \wrt nominal solution under all possible realizations of the uncertainty, but we note that other formulations are possible, \eg to optimize for the best-case realization ($\|\mathcal J(\ve P_k) - \min_{\ve p\in\ve P_k} \mathcal J(\ve p)\|_2^2$). We use a short-hand here for the interval-valued expressions, where $f(\ve P) :=\{f(\ve p) | \ve p\in\ve P\}$.
The objective in~\eqref{eq:prob_rob_nco} is of infinite-dimensional nature because of the set-valued expression $\mathcal J(\ve P_k)$. One possibility to remedy this situation is to model the uncertainty evolution through the dynamic system as a set of scenarios~\citep{luc13}, i.e., to take samples from the resulting sets, e.g. by using a multi-stage optimization approach. The problem~\eqref{eq:prob_rob_nco} should then be solved in a shrinking-horizon fashion at each sampling time to ensure the satisfaction of the end-point constraints. Another alternative exists in case full-state measurement is available. Then a feedback scheme~\citep{fra13} can be used to meet the terminal conditions.

\subsection{Implementation via dual robust control}
We adapt here the implicit dual-control methodology presented in~\cite{tha15, tha18} in this study. It models the evolution of the uncertainty in the states and parameters as a tree of discrete realizations of the uncertainty.
\begin{figure}
\psfrag{a1}{\hspace*{-1mm}\treetextsize { $\ve x_0$}}
\psfrag{b1}{\hspace*{-1mm}\treetextsize { $u_0^1,$}}
\psfrag{b2}{\hspace*{-1mm}\treetextsize { $\ \ve p_0^1$}}
\psfrag{b3}{\hspace*{-1mm}\treetextsize { $u_0^2,$}}
\psfrag{b4}{\hspace*{-1mm}\treetextsize { $\ \ve p_0^2$}}
\psfrag{b5}{\hspace*{-1mm}\treetextsize { $u_0^3,$}}
\psfrag{b6}{\hspace*{-1mm}\treetextsize { $\ \ve p_0^3$}}
\psfrag{c1}{\hspace*{-1.mm}\treetextsize { $\ve x_1^1$}}
\psfrag{c2}{\hspace*{-1.mm}\treetextsize { $\ve x_1^2$}}
\psfrag{c3}{\hspace*{-1.mm}\treetextsize { $\ve x_1^3$}}
\psfrag{d1}{\hspace*{-1mm}\treetextsize { $u_1^1,$}}
\psfrag{d2}{\hspace*{-1mm}\treetextsize { $\,\ \ve p_1^1$}}
\psfrag{d3}{\hspace*{-1mm}\treetextsize { $u_1^2,$}}
\psfrag{d4}{\hspace*{-1mm}\treetextsize { $\ \ve p_1^2$}}
\psfrag{d5}{\hspace*{-1mm}\treetextsize { $u_1^3,$}}
\psfrag{d6}{\hspace*{-1mm}\treetextsize { $\ \ve p_1^3$}}
\psfrag{e1}{\hspace*{-1mm}\treetextsize { $u_1^4,$}}
\psfrag{e2}{\hspace*{-1mm}\treetextsize { $\ \ve p_1^4$}}
\psfrag{e3}{\hspace*{-1mm}\treetextsize { $u_1^5,$}}
\psfrag{e4}{\hspace*{-1mm}\treetextsize { $\ \ve p_1^5$}}
\psfrag{e5}{\hspace*{-1mm}\treetextsize { $u_1^6,$}}
\psfrag{e6}{\hspace*{-1mm}\treetextsize { $\ \ve p_1^6$}}
\psfrag{f1}{\hspace*{-1mm}\treetextsize { $u_1^7,$}}
\psfrag{f2}{\hspace*{-1mm}\treetextsize { $\,\ \ve p_1^7$}}
\psfrag{f3}{\hspace*{-1mm}\treetextsize { $u_1^8,$}}
\psfrag{f4}{\hspace*{-1mm}\treetextsize { $\ \ve p_1^8$}}
\psfrag{f5}{\hspace*{-1mm}\treetextsize { $u_1^9,$}}
\psfrag{f6}{\hspace*{-1mm}\treetextsize { $\ \ve p_1^9$}}
\psfrag{g1}{\hspace*{-1.mm}\treetextsize { $\ve x_2^1$}}
\psfrag{g2}{\hspace*{-1.mm}\treetextsize { $\ve x_2^2$}}
\psfrag{g3}{\hspace*{-1.mm}\treetextsize { $\ve x_2^3$}}
\psfrag{g4}{\hspace*{-1.mm}\treetextsize { $\ve x_2^4$}}
\psfrag{g5}{\hspace*{-1.mm}\treetextsize { $\ve x_2^5$}}
\psfrag{g6}{\hspace*{-1.mm}\treetextsize { $\ve x_2^6$}}
\psfrag{g7}{\hspace*{-1.mm}\treetextsize { $\ve x_2^7$}}
\psfrag{g8}{\hspace*{-1.mm}\treetextsize { $\ve x_2^8$}}
\psfrag{g9}{\hspace*{-1.mm}\treetextsize { $\ve x_2^9$}}
\psfrag{h1}{\hspace*{-1mm}\treetextsize { $u_2^1,$}}
\psfrag{h2}{\hspace*{-1mm}\treetextsize { $\,\ \ve p_2^1$}}
\psfrag{h3}{\hspace*{-1mm}\treetextsize { $u_2^2,$}}
\psfrag{h4}{\hspace*{-1mm}\treetextsize { $\,\ \ve p_2^2$}}
\psfrag{h5}{\hspace*{-1mm}\treetextsize { $u_2^3,$}}
\psfrag{h6}{\hspace*{-1mm}\treetextsize { $\,\ \ve p_2^3$}}
\psfrag{i1}{\hspace*{-1mm}\treetextsize { $u_2^4,$}}
\psfrag{i2}{\hspace*{-1mm}\treetextsize { $\,\ \ve p_2^4$}}
\psfrag{i3}{\hspace*{-1mm}\treetextsize { $u_2^5,$}}
\psfrag{i4}{\hspace*{-1mm}\treetextsize { $\,\ \ve p_2^5$}}
\psfrag{i5}{\hspace*{-1mm}\treetextsize { $u_2^6,$}}
\psfrag{i6}{\hspace*{-1mm}\treetextsize { $\,\ \ve p_2^6$}}
\psfrag{j1}{\hspace*{-1mm}\treetextsize { $u_2^7,$}}
\psfrag{j2}{\hspace*{-1mm}\treetextsize { $\,\ \ve p_2^7$}}
\psfrag{j3}{\hspace*{-1mm}\treetextsize { $u_2^8,$}}
\psfrag{j4}{\hspace*{-1mm}\treetextsize { $\,\ \ve p_2^8$}}
\psfrag{j5}{\hspace*{-1mm}\treetextsize { $u_2^9,$}}
\psfrag{j6}{\hspace*{-1mm}\treetextsize { $\,\ \ve p_2^9$}}
\psfrag{k1}{\hspace*{-1.mm}\treetextsize { $\ve x_3^1$}}
\psfrag{k2}{\hspace*{-1.mm}\treetextsize { $\ve x_3^2$}}
\psfrag{k3}{\hspace*{-1.mm}\treetextsize { $\ve x_3^3$}}
\psfrag{k4}{\hspace*{-1.mm}\treetextsize { $\ve x_3^4$}}
\psfrag{k5}{\hspace*{-1.mm}\treetextsize { $\ve x_3^5$}}
\psfrag{k6}{\hspace*{-1.mm}\treetextsize { $\ve x_3^6$}}
\psfrag{k7}{\hspace*{-1.mm}\treetextsize { $\ve x_3^7$}}
\psfrag{k8}{\hspace*{-1.mm}\treetextsize { $\ve x_3^8$}}
\psfrag{k9}{\hspace*{-1.mm}\treetextsize { $\ve x_3^9$}}
\psfrag{l1}{\hspace*{-1mm}\treetextsize { $u^*(\ve p_2^1)$}}
\psfrag{l2}{\hspace*{-1mm}\treetextsize { }}
\psfrag{l3}{\hspace*{-1mm}\treetextsize { $u^*(\ve p_2^2)$}}
\psfrag{l4}{\hspace*{-1mm}\treetextsize { }}
\psfrag{l5}{\hspace*{-1mm}\treetextsize { $u^*(\ve p_2^3)$}}
\psfrag{l6}{\hspace*{-1mm}\treetextsize { }}
\psfrag{m1}{\hspace*{-1mm}\treetextsize { $u^*(\ve p_2^4)$}}
\psfrag{m2}{\hspace*{-1mm}\treetextsize { }}
\psfrag{m3}{\hspace*{-1mm}\treetextsize { $u^*(\ve p_2^5)$}}
\psfrag{m4}{\hspace*{-1mm}\treetextsize { }}
\psfrag{m5}{\hspace*{-1mm}\treetextsize { $u^*(\ve p_2^6)$}}
\psfrag{m6}{\hspace*{-1mm}\treetextsize { }}
\psfrag{n1}{\hspace*{-1mm}\treetextsize { $u^*(\ve p_2^7)$}}
\psfrag{n2}{\hspace*{-1mm}\treetextsize { }}
\psfrag{n3}{\hspace*{-1mm}\treetextsize { $u^*(\ve p_2^8)$}}
\psfrag{n4}{\hspace*{-1mm}\treetextsize { }}
\psfrag{n5}{\hspace*{-1mm}\treetextsize { $u^*(\ve p_2^9)$}}
\psfrag{n6}{\hspace*{-1mm}\treetextsize { }}
\psfrag{o1}{\hspace*{-1.mm}\treetextsize { $\ve x^1(t_\ff^1)$}}
\psfrag{o2}{\hspace*{-1.mm}\treetextsize { $\ve x^2(t_\ff^2)$}}
\psfrag{o3}{\hspace*{-1.mm}\treetextsize { $\ve x^3(t_\ff^3)$}}
\psfrag{o4}{\hspace*{-1.mm}\treetextsize { $\ve x^4(t_\ff^4)$}}
\psfrag{o5}{\hspace*{-1.mm}\treetextsize { $\ve x^5(t_\ff^5)$}}
\psfrag{o6}{\hspace*{-1.mm}\treetextsize { $\ve x^6(t_\ff^6)$}}
\psfrag{o7}{\hspace*{-1.mm}\treetextsize { $\ve x^7(t_\ff^7)$}}
\psfrag{o8}{\hspace*{-1.mm}\treetextsize { $\ve x^8(t_\ff^8)$}}
\psfrag{o9}{\hspace*{-1.mm}\treetextsize { $\ve x^9(t_\ff^9)$}}
\psfrag{a2}{\hspace*{-1mm}\small{}}
\psfrag{a3}{\hspace*{-1mm}\small{Robust horizon $N_r = 2$}}
\centering
\includegraphics[width=0.7\textwidth]{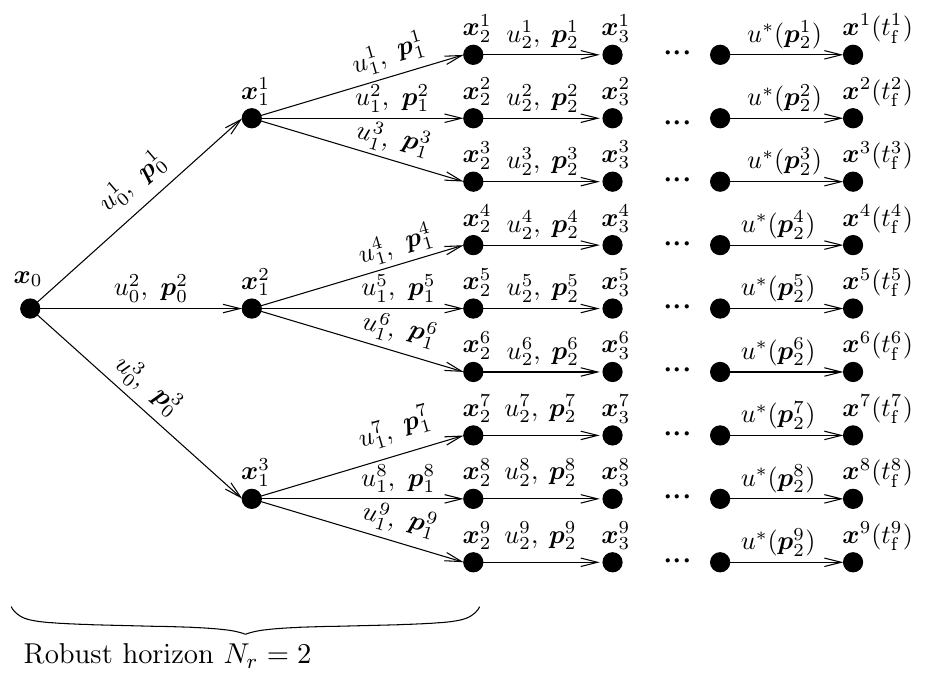} 
\caption{Scenario tree representation of the uncertainty evolution for a
multi-stage controller.}
\label{fig:rn:robusttree} 
\end{figure}
\begin{subequations}\label{eq:objective_dual}
\begin{align}
&\min_{\stackrel{t_1^j, t_2^j, t_\ff^j, \forall j\in I}
{u^{j}(t_k), \forall k\leq N_r, \forall j\in I}} \
\|\tilde{\mathcal J}(\ve p) - \mathcal J(\ve p_0)\|_2^2\\
 \st &\ \forall j\in I:\notag\\
 &\ \dot{\ve x}(t, \ve p_k^{r(j)}) = \ve f_0(\ve x(t, \ve p_k^{r(j)}), \ve p_k^{r(j)}) + \ve f_u(\ve x(t, \ve p_k^{r(j)}), \ve p_k^{r(j)}) u^j(t_k), \ \forall k<N_r, \\
 &\ \dot{\ve x}(t, \ve p_k^{r(j)}) = \ve f_0(\ve x(t, \ve p_k^{r(j)}), \ve p_k^{r(j)}) + 
 \ve f_u(\ve x(t, \ve p_k^{r(j)}), \ve p_k^{r(j)}) u^*(t_k, \ve \pi^{r(j)}), \ \forall k\geq N_r,\\
 &u^j(t_k)=u^l(t_k)\; \text{if } \ve x_k(t_k, \ve p_k^{r(j)}) = \ve x(t_k, \ve p_k^{r(l)}), 
 \ \forall l \in I, \ \forall k<N_r,\label{eq:nonanticip}\\
 &\ve p_{k+1}^{r(j)}= h(\ve x(t_k, \ve p_k^{r(j)}), u^j(t_k), \ve p_k^{r(j)}),
 \ \forall k<N_r,\\
 & \ve x^j(t_\ff, \ve p_k^{r(j)}) = \ve x_\ff,
\end{align}
\end{subequations}
where $\tilde{\mathcal J}(\ve P_k) := (\mathcal J(\ve p_{N_r}^{r(1)}),
\mathcal J(\ve p_{N_r}^{r(2)}), \dots, \mathcal J(\ve p_{N_r}^{r(n_s)}))^T$.
We adopt the notation from~\cite{lucia2014b, jan16} where index $k$ denotes the sample-and-hold value of a variable on the interval $[t_k, t_{k+1}]$, $j$
represents a particular realization of uncertainty and $p(j)$ is the realization at the parent node of the scenario tree (Fig.~\ref{fig:rn:robusttree}). The tree contains $n_s$ scenarios that correspond to the index set $I$ of the uncertainty propagation through dynamics of the system. The function $h(\cdot)$ denotes the estimation procedure~\eqref{eq:nlp_par_bounds}. The value of $N_r$ represents the length of the so-called robust horizon, which marks the stage, until which the tree is considered to branch. Note that this models a possible variability in the parametric uncertainty and, in proposed methodology, it models the estimation of the bounds of uncertain parameters. Note also that the control inputs are free until the stage $N_r$---they only need to fulfill the non-anticipativity constraints~\eqref{eq:nonanticip}---so the proposed scheme shows a significant reduction of the number of degrees of freedom of the optimization as opposed to the situation, where only the multi-stage approach (equivalent under some assumptions to robust dynamic programming~\cite{lee97}) would be used without the parameterized solution to nominal optimal control problem. The value of $N_r$ should be set as large as possible, ideally until the stage when the earliest possible switching of the optimal control input occurs. A similar approach is utilized for uncertainty propagation in set-membership context~\cite{you17}. However, as the simulation experiments have shown for standard multi-stage predictive control~\cite{luc13}, $N_r=1$ or $N_r=2$ is a practical and sufficient choice \wrt to the performance of the scheme in most cases.

A possible interpretation of the presented dual-control scheme is that:
\begin{itemize}
 \item the optimal excitation of the system, which results in improved precision  of parameter bounds, is obtained as a consequence of minimization of the variance of the objective function under uncertainty and by freeing the (initial) control moves on the robust horizon from the optimality conditions of~\eqref{eq:prob_gen};
 \item the optimality of each scenario is guaranteed beyond $N_r$ by the control parameterization using optimality conditions and from the principle of dynamic programming, which means that, despite initial control moves are not fixed, the control moves until the end of the horizon are optimal \wrt state values of each scenarios.
\end{itemize}

Regarding the computational aspect, the presented dual-control scheme is of the same complexity as the multi-stage NMPC with (short) prediction horizon set to $N_r$, which clearly shows the computational benefits.

The real-time implementation of the proposed scheme proceeds as shown
in Algorithm~\ref{alg:dual}. Herein, $T_s$ represents sampling time of
the plant, and $\text{mid}(\cdot)$ and $\text{diam}(\cdot)$ stand for
mid-point and diameter of an interval, respectively. User-specified tolerance $\varepsilon$ can be used for a further speed-up of online computations. Its use is motivated by avoiding of re-calculation of the control profile if the optimal switching times are known with sufficient accuracy (e.g., the width of uncertainty in switching time $t_i$ might be less than the sampling period).

\begin{algorithm}
\caption{Real-time implementation of dual robust controller.}
\label{alg:dual}
\begin{flushleft}
 \textbf{Require:} $\ve P_0, \varepsilon>0, T_s, N_r$\\
\textbf{Initialization:} Calculate nominal $u^\ast(t, \ve\pi)$ for $\ve x_0$ and $\text{mid}(\ve P_0)$ and discretize the profile according to the sampling time $T_s$ to get $u^\ast(k, \ve\pi), \forall k\geq0$ . Evaluate $[t_i], \ \forall t_i\in\{t_1, t_2, t_\ff\}$. Set $t:=0$.\\
\textbf{Main Loop:}
\begin{itemize}
\item[] \textbf{For} $t_i\in\{t_1, t_2, t_\ff\}$
\begin{itemize}
\item[] \textbf{While} $t\leq \text{mid}([t_i])-T_s$
\begin{itemize}
\item[] \textbf{If} $\text{diam}([t_i])\geq\varepsilon$
\begin{itemize}
\item[] $u^\ast(k, \ve\pi),\, \forall k\geq t/T_s \ \leftarrow$ Solve~\eqref{eq:objective_dual} with $\ve P_{t/T_s}$ and discretize the obtained profile according to $T_s$.
\end{itemize}
\item[] \textbf{End If}
\item[] Apply $u^\ast(t/T_s)$ to the plant.
\item[] Obtain new measurements and increment $t:=t+T_s$.
\item[] $\ve P_{t/T_s}\ \leftarrow$ Solve~\eqref{eq:nlp_par_bounds}.
\item[] Update $[t_i], \ \forall t_i\in\{t_1, t_2, t_\ff\}$.
\end{itemize}
\item[] \textbf{End While}
\end{itemize}
\item[] \textbf{End For}
\end{itemize}
\end{flushleft}
\end{algorithm}

\subsection{Possible extensions}
Several extensions of the proposed scheme might be foreseen at this stage. These mostly depend on the actual problem at hand and its complexity.
\begin{description}
 \item [Handling discontinuities in the control strategies]
 Note that because of the switching nature of the optimal control strategy, the proposed problem might show discontinuity as a consequence of activation of the input constraints based on uncertain parameters ($S(\ve x(t), \ve p)$). Simply speaking, it may happen that for a subset of $\ve P$ the resulting optimal sequence commences with $u(t)=u^L$ and vice versa for other subset of $\ve P$. This can be remedied by an adaptation of the continuous-formulation
 technique for scheduling~\cite{pra11}.
 \item [Handling the complexity of the estimation problem]
 Clearly, the presented strategy---having the estimation problem embedded in the constraints---is computationally tractable for only specific estimation problems, e.g., when mathematical model is linear in parameters. Here either a strategy based on approximate linear (linearization-based) estimation can be used or an approach that estimates contribution of each measurement based on parametric sensitivities~\citep{tha15}.
 \item [Handling the complexity of the optimization problem]
 Despite the approach presented in this section reduces the complexity of implementation of a model-predictive controller that needs to optimize all the control moves from the initial to the final time point, situations still exist where a combination of long time horizon and model complexity may result in an intractable optimization problem~\eqref{eq:objective_dual}. In this case, approximate dynamic programming techniques~\citep{LEE20051281} enhanced with the knowledge of optimal control structure might offer a viable alternative.
\end{description}

\section{Case study}
\begin{figure}
\centering
\psfrag{Feed}[b][c][1][0]{feed tank}
\psfrag{alpha}[c][t][1][0]{$\quad u(t)$}
\psfrag{qo}[b][c][1][0]{}
\psfrag{qp}[b][c][1][0]{$q_p$} 
\psfrag{NF}[b][c][1][0]{nanofiltration}
\psfrag{mem}[b][c][1][0]{membrane}
\psfrag{coolant}[b][c][1][0]{coolant}
\psfrag{canal}[b][c][1][0]{canal}
\psfrag{pump}[b][c][1][0]{pump}
\psfrag{Permeate}[b][c][1][0]{   permeate}
\psfrag{permeate}[b][c][1][0]{permeate}
\psfrag{tank}[b][c][1][0]{tank}
\psfrag{retentate}[b][c][1][0]{retentate}
\includegraphics[width=0.7\textwidth]{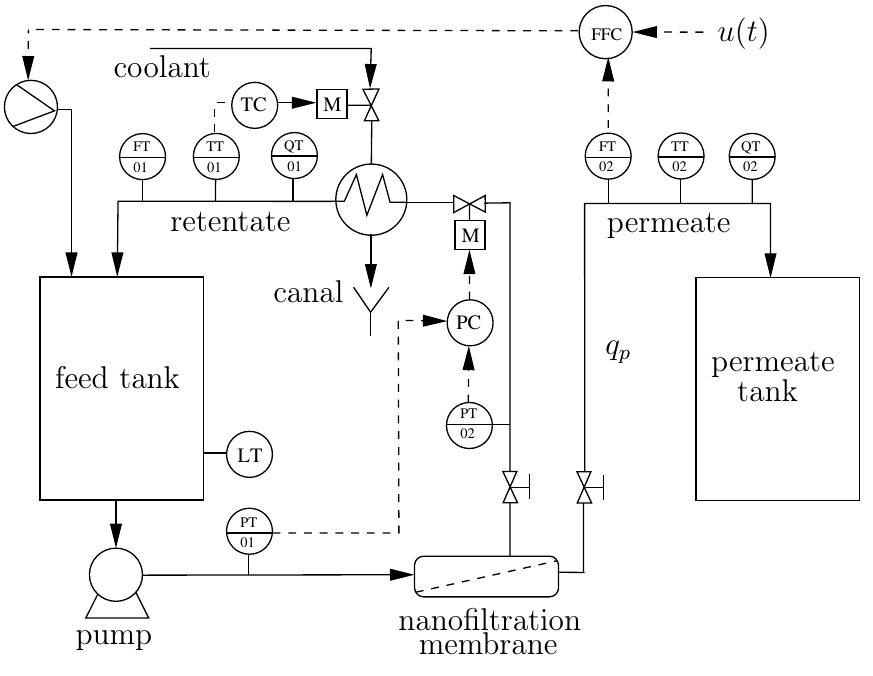}
\caption{Nanodiafiltration process scheme.}
\label{fig:plant_scheme}
\end{figure}
We consider a case study of time-optimal control of a batch diafiltration
process from~\cite{pau12jms}. The scheme of the plant is shown in
Fig.~\ref{fig:plant_scheme}. The goal is to process a solution with initial
volume ($V_0$) that is fed into the feed tank at the start of the batch and
that comprises two solutes of initial concentrations $c_{1,0}$ and $c_{2,0}$. At the end of the batch, the prescribed final concentrations $c_{1,\ff}$ and
$c_{2,\ff}$ must be met. The transmembrane pressure is controlled at a constant value. The temperature of the solution is maintained around a constant value using a heat exchanger. The manipulated variable $u(t)$ is the ratio between fresh water inflow into the tank and the permeate outflow $q_p$ that is given by
\begin{align}\label{eq:ext_lim_model}\notag
q_p = \gamma_1 \ln\left(\frac{\gamma_2}{c_1 c_2^{\gamma_3}}\right) &=
     \gamma_1\left(\ln(\gamma_2) - \ln(c_1) - \gamma_3\,\ln(c_2)\right),\\
     &=p_1 - p_2\ln(c_1(t)) - p_3\ln(c_2(t)),
\end{align}
where the model parameterized with $\gamma_1$, $\gamma_2$, and $\gamma_3$ offers phenomenological interpretation of the parameters while the parameterization using $p_1$, $p_2$, and $p_3$ gives a model more appropriate for parameter estimation. The permeate is measured at intervals of one minute with the assumed measurement noise that is bounded by $\sigma = 1\times10^{-2}\unit{L/h}$. The model of the permeate flux can be reduced to another widely used \emph{limiting flux model} if $\gamma_3 = 0$, so this case study offers to study both parametric and non-parametric plant-model mismatch. 

Concentrations of both components $c_1(t)$ and $c_2(t)$, where the first
component is retained by the membrane and the second one can freely pass
through, are measured as well and will be assumed to be perfectly known. This is only assumed for simplicity as the resulting estimation problem is of a static nature. Should an uncertainty be considered in measured values of $c_1(t)$ and $c_2(t)$, an error-in-variables approach~\citep{sod07} can be adopted for parameter estimation.

The objective is to find a time-dependent input function $u(t)$, which
guarantees the transition from the given initial $c_{1,0}, c_{2,0}$ to final
$c_{1,\ff}, c_{2,\ff}$ concentrations in minimum time. This problem can be
formulated as:
\begin{subequations}\label{eq:t_opt_prob}
\begin{align}\label{eq:obj_func}
\mathcal{J}^{\ast} &= \min_{u(t)\in [0,\infty)}  \int_{0}^{t_\ff}\!1\,\dt,\\
\st \ \dot c_1&=\frac{c_1^2q_p}{c_{1,0}V_0}(1-u), \
c_1(0)=c_{1,0},\ c_1(t_{\ff}) =c_{1,\ff} \label{eq:c1}\\
\dot c_2&= - \frac{c_1 c_2q_{\pp}}{c_{1,0}V_0}u, \
c_2(0)=c_{2,0}, \ c_2(t_{\ff}) =c_{2,\ff}\label{eq:c2}\\
q_p &= \gamma_1\left(\ln(\gamma_2) - \ln(c_1) - \gamma_3\,\ln(c_2)\right).
\end{align}
\end{subequations}
The (nominal) parameters of the problem are $c_{1,0} = 50\,\unit{g/L}$,
$c_{1,\ff} = 150\,\unit{g/L}$, $c_{2,0} = 50\,\unit{g/L}$,
$c_{2,\ff} = 0.05\,\unit{g/L}$, $V_0  = 20\,\unit{L}$,
$\gamma_1 = 3\,\unit{L/h}$,
$\gamma_2 = 1000\,\unit{g/L}$, $\gamma_3 = 0.1$. The extremal values of
$u(t)$ stand for a mode with no water addition, \ie pure filtration, when $u(t)=0$ and pure dilution, \ie a certain amount of water is added at a single time instant, $u(t)=\infty$.

As the problem involves end-point constraints, the constraint satisfaction must be ensured using the shrinking-horizon strategy. The arising real-time optimization problem is thus computationally demanding even in its nominal setup. 

The nominal (parameterized) optimal control of this process can be identified
using Pontryagin's minimum principle~\citep{pon62} as:
\begin{equation}\label{eq:singular_ctrl_switch_cs}
u^\ast(t, \ve\pi) =
\begin{cases}
  0, &  t\in[0, t_1], \ S(\ve x(t, \ve p), \ve p) >0,\\
  \infty, & t\in[0, t_1], \ S(\ve x(t, \ve p), \ve p) <0,\\
  u_s, & t\in[t_1, t_2], \ S(\ve x(t, \ve p), \ve p) =0,\\
  0, & t\in[t_2,t_\ff], \ S(\ve x_\ff, \ve p) <0,\\
  \infty, & t\in[t_2,t_\ff], \ S(\ve x_\ff, \ve p) >0,
\end{cases}
\end{equation}
where the singular control and the respective switching function can be found
explicitly~\citep{pau12jms} as
\begin{align}\label{eq:singular_ctrl_switch_cs2}
 u_s(\ve x(t, \ve p), \ve p) &:= \frac{1}{1+\gamma_3},\\
 S(\ve x(t, \ve p), \ve p) &:= \gamma_1\left(\ln(\gamma_2) - \ln(c_1)
 - \gamma_3\,\ln(c_2) - \gamma_3 - 1\right).
\end{align}

We consider the parametric uncertainty in $\gamma_1$, $\gamma_2$, and $\gamma_3$ to be given by $\pm10\%$ \wrt the nominal values. In simulation studies, the true values of the parameters are chosen randomly from this range. It is clear (from~\eqref{eq:singular_ctrl_switch_cs2}) that the real-time optimality of the operation is strongly influenced by the accuracy of the estimation of the parameters $\gamma_2$ and $\gamma_3$. Preliminary numerical tests with optimal experiment design (OED) methodology~\citep{mupa_cace17} showed that for the most accurate estimation of $\gamma_2$ the manipulated variable $u(t)=0$ and, on the other hand, the best estimation accuracy of $\gamma_3$ is reached when $u(t)=1$. This shows a mutual benefit of the optimal control strategy being a sequence $u^* = \{0, \approx 1, \infty\}$ and estimation of $\gamma_2$, and a potential conflict of accurate estimation of $\gamma_3$ and the optimal control policy. This can also be seen from~\eqref{eq:ext_lim_model} and~\eqref{eq:c2}, where it is clear that when the (nominally) optimal controller applies $u(t)=0$, the parameter $\gamma_3$ is unidentifiable as the concentration $c_2(t)$ remains constant. The OED studies also showed that the best time to excite the plant is in the beginning of the operation. This stems from the absolute error of the measurement (see~\eqref{eq:bound_meas2}) and from the fact that the measured permeate flux is highest in the beginning of the operation and drops dramatically with the increase of concentration $c_1(t)$.

The optimal controller, i.e., one that possesses the information about true parameter values achieves the performance $t_\ff = 9.25\,\unit{h}$. Under the governance of the robust adaptive controller, the batch is finished $9.27$ hours. This small difference, despite the adaptive controller does not excite the plant optimally, comes as a consequence of the constant singular control, the nature of the singular control in general ($\partial H/\partial u=0$, see the discussion in~\cite{sri03a}), and the ability of the controller to guess the value of $t_1$ relatively well despite the imprecise estimates.

As expected, the adaptive controller follows the control profile of the nominally optimal scheme, i.e., $(u_1, u_2, u_3)^T=(0, 0, 0)^T$, while adapting the switching times and the value of singular control as new information becomes available. The dual controller excites the plant in the beginning of the operation by choosing the control profile $u=(0, 1, 0, 0, \dots)^T$ instead of the operation with $u(t)=0, \forall t\leq t_1$. This allows for a more precise estimation of the parameters and as a result the achieved performance is practically the same as the performance of the optimal controller.

\begin{figure}
\centering
\includegraphics[width=0.8\textwidth]{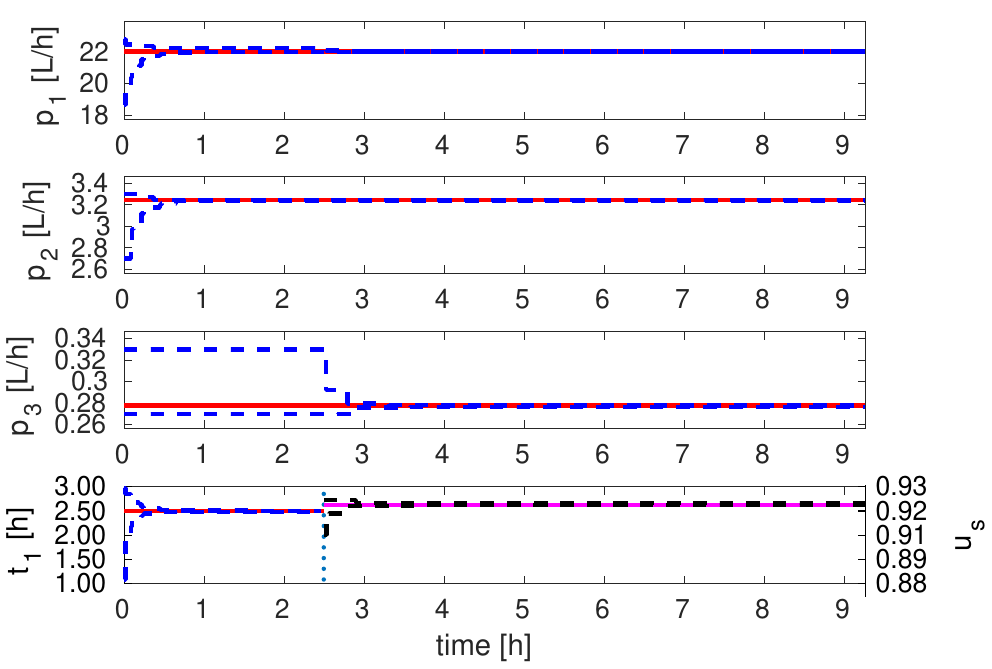}
\caption{Results obtained using the adaptive strategy. Set-membership estimates over time (top three plots) with projection of the uncertainty in the parameters on the switching time $t_1$ (left-hand part in the bottom plot) and on the value of $u_s$ (right-hand part in the bottom plot). The true (optimal) values are shown as solid lines, the bounds are represented using dashed lines. The vertical line in the bottom plot indicates the optimal switching time.}
\label{fig:perf_est}
\end{figure}

\begin{figure}
\centering
\includegraphics[width=0.8\textwidth]{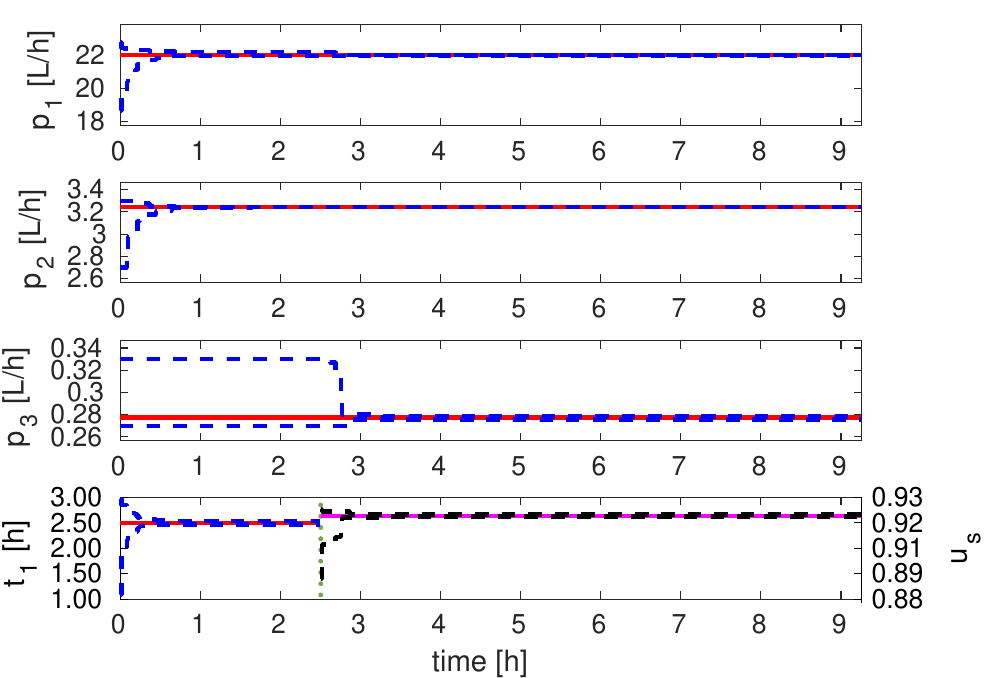}
\caption{Results obtained using the dual-control strategy. Set-membership estimates over time (top three plots) with projection of the uncertainty in the parameters on the switching time $t_1$ (left-hand part in the bottom plot) and on the value of $u_s$ (right-hand part in the bottom plot). The true (optimal) values are shown as solid lines, the bounds are represented using dashed lines. The vertical line in the bottom plot indicates the optimal switching time.}
\label{fig:perf_est_dual}
\end{figure}

Figures~\ref{fig:perf_est} and~\ref{fig:perf_est_dual} present performance of the estimation (in terms of estimated parameter bounds) throughout the run of the batch for adaptive and dual controller ($N_r=1$). It is clear that the bounds on both parameters are dramatically reduced around the time point of 0.5\,h, which precedes the time point $t_1^\text{opt}$, when the switch in the control input should be executed. The bottom plots in Figs.~\ref{fig:perf_est} and~\ref{fig:perf_est_dual} also show the evolution of the uncertainty in $t_1$, which is projected using interval-based calculations, and of the uncertainty in the singular control $u_s$. It should be noted here that the both the approaches are successful here mainly since the applied control input in the first arc coincides with an input that would result from a dynamic optimal-experiment design study. Here, $u(t)=0$ ensures the fastest possible increase of concentration $c_1(t)$, which reveals the most informative measurements about $\gamma_2$.

\begin{figure}
\psfrag{strategy}[t][c][1][0]{strategy}
\psfrag{performance}[b][c][1][0]{batch time\,[h]}
\centering
\includegraphics[width=0.6\textwidth]{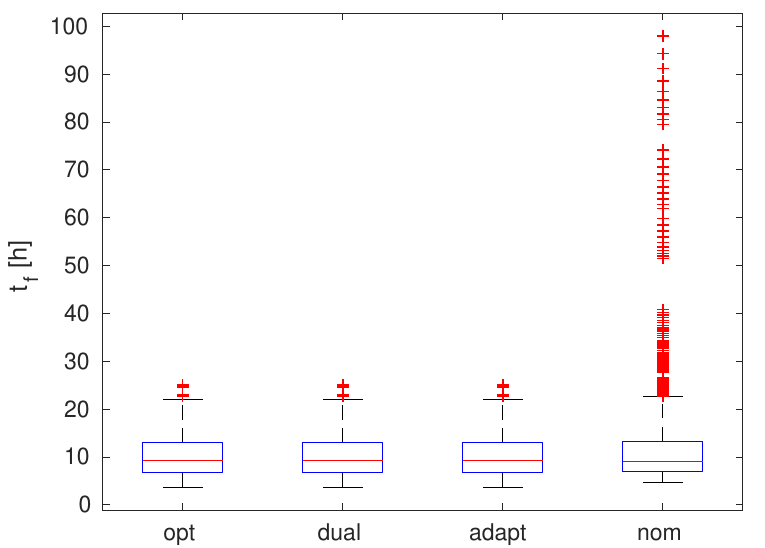}
\caption{A box plot with the statistical information (the median as solid horizontal line, the 25$^\text{th}$ and 75$^\text{th}$ percentiles as boxes, statistical extremes as ends of the whiskers, and the outliers as `\textsf{+}') about the performance of the different control strategies.}
\label{fig:perf_comp}
\end{figure}

Figure~\ref{fig:perf_comp} shows the box plot statistics of the
performance of the studied controllers in 1,000 simulations with
different true values of parameters $\ve p$ taken from uniform grid of
$\ve P_0$. The plot shows the median, the 25$^\text{th}$ and 75$^\text{th}$ percentiles and the outliers.
It is clear that adaptive and dual controller reach
performance very close to the optimal one. They also greatly reduce
the variance of the nominal controller. It is also clear for this case study that for the actual realization of the control of this plant, a dual controller would not be essential. So in this case, the presented methodology would serve in the design phase to assess the need of advanced robust adaptive controller.

Regarding the computational aspects, the robust adaptive, the nominal (and the optimal) controllers can be resolved analytically and thus require only an evaluation of the parameterized control law at each sampling in the plant. Thus they are applicable in real time. The dual controller then clearly requires much higher computational effort. This is in order of minutes in our current naive implementation. Due to the nature of the treated case study, it is, however, possible to determine the optimal dual control profile off-line as the probing action of the controller is realized in the first two sampling intervals.

\section{Conclusion}
We have presented a novel methodology for dual robust controller
design for the (real-time) optimal control of batch processes. The
controller achieves the dual action by direct consideration of the
effects of the future exciting control signal on the performance of
the plant. The crucial step is the parameterization of the (open-loop)
optimal controller. This allows for adaptation and implementation of
the dual robust control strategies devised earlier in the
literature. The benefits of the approach were shown in the case study
on batch membrane filtration. Set-membership estimation was used for
the parameter estimation, as a technique that can provide guaranteed
bounds on the parametric uncertainty. The future work will concentrate
on the experimental validation of the presented methodology at the
laboratory membrane plant.

\begin{acknowledgement}
The authors acknowledge Sakthi Thangavel from TU Dortmund for creating the graphical scheme of the multi-stage approach. We are also grateful for the constructive comments of the anonymous reviewers. We gratefully acknowledge the contribution of the Scientific Grant Agency of the Slovak Republic under the grant 1/0004/17, of the Slovak Research and Development Agency under the project APVV 15-0007 and of the European Commission under the grant 790017 (GuEst). This publication is also a partial result of the Research \& Development Operational Programme for the project University Scientific Park STU in Bratislava, ITMS 26240220084, supported by the Research 7 Development Operational Programme funded by the ERDF.
\end{acknowledgement} 

\appendix
\section{Appendix}
\subsection{Conditions for Optimality}
Pontryagin's minimum principle can be used~\citep{sri03a} to identify the optimal solution to~\eqref{eq:prob_gen} via enforcing the necessary conditions for minimization of a Hamilton function (Hamiltonian)
\begin{align}\label{eq:hamiltonian}
 H:= \mu^L(u^L-u) + \mu^U(u-u^U) + \underbrace{F_0
 +\ve\lambda^T\ve f_0}_{H_0(\ve x(t, \ve p), \ve \lambda(t, \ve p), \ve p)} \!+ \underbrace{\left(F_u+\ve\lambda^T\ve f_u \right)}
 _{H_u(\ve x(t, \ve p), \ve \lambda(t, \ve p), \ve p)}u,
\end{align}
where $\ve\lambda(t)$ is a vector of adjoint variables, which are defined through
\begin{align}
&\dot{\ve\lambda}(t, \ve p)=-\frac{\partial H}{\partial \ve x}(t, \ve p),
\quad \ve\lambda(t_\ff, \ve p)=\ve\nu(\ve p),
\end{align}
and $\mu^L(t,\ve p)$, $\mu^U(t,\ve p)$, and $\ve\nu(\ve p)$ are the Lagrange multipliers associated with bounds on control input and end-point constraints.
The minimization is carried out such that
\begin{subequations}\label{eq:pmp}
\begin{align}
 &\min_{u(t), \ve\nu, t_\ff, \mu^L(t)\geq0, \mu^U(t)\geq0}\ H(\ve x(t, \ve p), \ve\lambda(t, \ve p), u(t), \ve p, \ve\nu, t_\ff, \mu^L(t), \mu^U(t))\\
 \text{s.t. } &\dot{\ve x}(t, \ve p)=\ve f_0 (\ve x(t, \ve p), \ve p) + \ve f_u(\ve x(t, \ve p), \ve p) u(t), \quad
\ve x(0)=\ve x_0,\quad \ve x(t_\ff, \ve p)=\ve x_\ff,\\
&\dot{\ve\lambda}(t, \ve p)=-\frac{\partial H}{\partial \ve x}(t, \ve p),
\quad \ve\lambda(t_\ff, \ve p)=\ve\nu(\ve p),\\ \label{eq:kkt_u}
 & \mu^L(u^L-u(t))=0, \quad \mu^U(u(t)-u^U)=0.
\end{align}\end{subequations}

The necessary conditions for optimality of~\eqref{eq:pmp} can be stated
as~\citep{sri03a}: $\forall t\in[0, t_\ff]$
\begin{align}
 \frac{\partial H}{\partial u}:=
 H_u(\ve x(t, \ve p), \ve \lambda(t, \ve p),  \ve p) - \mu^L(t) + \mu^U(t) &= 0,\label{eq:optcon_u}\\ 
 H(\ve x(t, \ve p), \ve \lambda(t, \ve p), \ve p, u(t), \mu^L(t), \mu^U(t))&=0,\label{eq:optcon_Heq0}\\
 H_0(\ve x(t, \ve p), \ve \lambda(t, \ve p), \ve p)&=0,\label{eq:optcon_H0eq0}\\
 \ve x(t_\ff, \ve p) - \ve x_\ff &=0.
\end{align}
The condition~\eqref{eq:optcon_Heq0} arises from the transversality, since the
final time is free~\citep{pon62}, and from the fact that the optimal Hamiltonian is constant over the whole time horizon, as it is not an~explicit function of time. The condition~\eqref{eq:optcon_H0eq0} is the consequence of the former two. Since the Hamiltonian is affine in input variable
(see~\eqref{eq:hamiltonian}), the optimal trajectory of control variable is
either determined by active input constraints or it evolves inside the feasible region. Let us first consider the latter case.

Assume that for some point $t$ we have $H_u(\cdot)=0$ and $u^L<u(t)<u^U$. It follows from~\eqref{eq:optcon_u} that the optimal control maintains $H_u(\cdot)=0$. Such control is traditionally denoted as singular. Further properties of the singular arc, such as switching conditions or state-feedback control trajectory can be obtained by differentiation of $H_u(\cdot)$ with respect to time (sufficiently many times) and by requiring the time derivatives of $H_u(\cdot)$ to be zero.  The time derivatives of $H(\cdot)$ and $H_0(\cdot)$ are equal to zero as well. Earlier results on derivation of optimal control for input-affine dynamic systems~\citep{jon90,sri03a} suggest that it is possible to eliminate adjoint variables from the optimality conditions and thus arrive at analytical characterization of switching conditions and optimal control for singular and saturated-control arcs.

As the optimality conditions obtained by the differentiation \wrt time are
linear in the adjoint variables, the differentiation of $H_u$ (or $H_0$) can be carried out until it is possible to transform the obtained conditions to a pure state-dependent switching function $S(\ve x, \ve p)$. It is usually convenient to use a determinant of the coefficient matrix of the equation system $\ve A\ve\lambda=\ve0$ for this.

The singular control $u_\text s(\ve x(t), \ve p)$ is found from
\begin{align}\label{eq:singular_ctrl_gen}
 \frac{\dd S}{\dt}(\ve x(t, \ve p), \ve p) &=
 \frac{\partial S}{\partial \ve x^T}(\ve x(t, \ve p), \ve p)\frac{\dd \ve x}{\dt}(t, \ve p)\notag\\
 &=\frac{\partial S}{\partial \ve x^T}(\ve x(t, \ve p), \ve p)\left[\ve f_0(\ve x(t, \ve p), \ve p) +\ve f_u(\ve x(t, \ve p), \ve p) u_\text s\right] =0,
\end{align}
as
\begin{equation}\label{eq:singular_ctrl_gen2}
 u_\text s (\ve x(t), \ve p)= -\left(\frac{\partial S}{\partial \ve x^T}(\ve x(t, \ve p), \ve p)
 \ve f_0(\ve x(t, \ve p), \ve p)\right)\bigg/\left(
 \frac{\partial S}{\partial \ve x^T}(\ve x(t, \ve p), \ve p)\ve f_u(\ve x(t, \ve p), \ve p)\right).
\end{equation}

There exist cases when the switching function $S(\ve x(t, \ve p), \ve p)$ is unidentifiable by the aforementioned procedure since it might be impossible to eliminate adjoint variables from Eq.~\eqref{eq:optcon_u}. This depends on the dimensionality of the problem and on the problem structure. In such cases, the differentiation of $H_u$ (or $H_0$) is carried out until the manipulated variable appears explicitly in one of the optimality conditions. It is then possible to devise an expression for singular control that is independent of adjoint variables. This is done by reducing the adjoint-affine system to triangular form from which the unknown adjoint variables can be expressed as functions of state variables.

\subsection{Set-membership estimation}
In order to estimate the model parameters, we will make use of plant outputs (measurements), which are expressed as:
\begin{align}
 \hat{\ve y}(t) = \ve g(\ve x(t,\ve p), \ve p),
\end{align}
where $\ve g(\cdot)$ is a continuously differentiable vector function. We will assume that the true output of the plant $\ve y_p(t)$ is corrupted with a (sensor) noise that is bounded with a known magnitude $\ve\sigma$. Thus, the measured output $\ve y(t)$ is such that
\begin{align}\label{eq:bound_meas}
 |\ve y_p(t) - \ve y(t)|\leq\ve\sigma,
\end{align}
where the absolute value is understood component-wise. In turn, the set-membership constraints apply in the form:
\begin{align}\label{eq:bound_meas2}
 |\hat{\ve y}(t) - \ve y(t)|\leq\ve\sigma.
\end{align}

\providecommand{\latin}[1]{#1}
\makeatletter
\providecommand{\doi}
  {\begingroup\let\do\@makeother\dospecials
  \catcode`\{=1 \catcode`\}=2 \doi@aux}
\providecommand{\doi@aux}[1]{\endgroup\texttt{#1}}
\makeatother
\providecommand*\mcitethebibliography{\thebibliography}
\csname @ifundefined\endcsname{endmcitethebibliography}
  {\let\endmcitethebibliography\endthebibliography}{}


\begin{mcitethebibliography}{54}
\providecommand*\natexlab[1]{#1}
\providecommand*\mciteSetBstSublistMode[1]{}
\providecommand*\mciteSetBstMaxWidthForm[2]{}
\providecommand*\mciteBstWouldAddEndPuncttrue
  {\def\EndOfBibitem{\unskip.}}
\providecommand*\mciteBstWouldAddEndPunctfalse
  {\let\EndOfBibitem\relax}
\providecommand*\mciteSetBstMidEndSepPunct[3]{}
\providecommand*\mciteSetBstSublistLabelBeginEnd[3]{}
\providecommand*\EndOfBibitem{}
\mciteSetBstSublistMode{f}
\mciteSetBstMaxWidthForm{subitem}{(\alph{mcitesubitemcount})}
\mciteSetBstSublistLabelBeginEnd
  {\mcitemaxwidthsubitemform\space}
  {\relax}
  {\relax}

\bibitem[Bock and Plitt(1984)Bock, and Plitt]{boc84}
Bock,~H.~G.; Plitt,~K.~J. {A multiple shooting algorithm for direct solution of
  optimal control problems}. \emph{Proceedings 9th IFAC World Congress
  Budapest} \textbf{1984}, \emph{XLII}, 243--247\relax
\mciteBstWouldAddEndPuncttrue
\mciteSetBstMidEndSepPunct{\mcitedefaultmidpunct}
{\mcitedefaultendpunct}{\mcitedefaultseppunct}\relax
\EndOfBibitem
\bibitem[Cuthrell and Biegler(1989)Cuthrell, and Biegler]{cut89}
Cuthrell,~J.~E.; Biegler,~L.~T. {Simultaneous Optimization and Solution Methods
  for Batch Reactor Control Profiles}. \emph{Computers \& Chemical Engineering}
  \textbf{1989}, \emph{13}, 49--62\relax
\mciteBstWouldAddEndPuncttrue
\mciteSetBstMidEndSepPunct{\mcitedefaultmidpunct}
{\mcitedefaultendpunct}{\mcitedefaultseppunct}\relax
\EndOfBibitem
\bibitem[Vassiliadis \latin{et~al.}(1994)Vassiliadis, Sargent, and
  Pantelides]{vas94a}
Vassiliadis,~V.~S.; Sargent,~R. W.~H.; Pantelides,~C.~C. {Solution of a Class
  of Multistage Dynamic Optimization Problems. 1. Problems without Path
  Constraints}. \emph{Industrial \& Engineering Chemistry Research}
  \textbf{1994}, \emph{33}, 2111--2122\relax
\mciteBstWouldAddEndPuncttrue
\mciteSetBstMidEndSepPunct{\mcitedefaultmidpunct}
{\mcitedefaultendpunct}{\mcitedefaultseppunct}\relax
\EndOfBibitem
\bibitem[Barton and Pantelides(1993)Barton, and Pantelides]{bar93}
Barton,~P.~I.; Pantelides,~C.~C. {gPROMS} - a Combined Discrete/Continuous
  Modelling Environment for Chemical Processing Systems. \emph{Simulation
  Series} \textbf{1993}, \emph{25}, 25--34\relax
\mciteBstWouldAddEndPuncttrue
\mciteSetBstMidEndSepPunct{\mcitedefaultmidpunct}
{\mcitedefaultendpunct}{\mcitedefaultseppunct}\relax
\EndOfBibitem
\bibitem[{Process Systems Enterprise}(1997--2009)]{gproms}
{Process Systems Enterprise}, gPROMS. 1997--2009\relax
\mciteBstWouldAddEndPuncttrue
\mciteSetBstMidEndSepPunct{\mcitedefaultmidpunct}
{\mcitedefaultendpunct}{\mcitedefaultseppunct}\relax
\EndOfBibitem
\bibitem[{\v{C}}i\v{z}niar \latin{et~al.}(2006){\v{C}}i\v{z}niar, Fikar, and
  Latifi]{dynopt}
{\v{C}}i\v{z}niar,~M.; Fikar,~M.; Latifi,~M.~A. {MATLAB} Dynamic Optimisation
  Code DYNOPT. {U}ser's Guide. 2006; \url{https://bitbucket.org/dynopt}\relax
\mciteBstWouldAddEndPuncttrue
\mciteSetBstMidEndSepPunct{\mcitedefaultmidpunct}
{\mcitedefaultendpunct}{\mcitedefaultseppunct}\relax
\EndOfBibitem
\bibitem[Houska \latin{et~al.}(2011)Houska, Ferreau, and Diehl]{Houska2011a}
Houska,~B.; Ferreau,~H.; Diehl,~M. {ACADO} {T}oolkit -- {A}n {O}pen {S}ource
  {F}ramework for {A}utomatic {C}ontrol and {D}ynamic {O}ptimization.
  \emph{Optimal Control Applications and Methods} \textbf{2011}, \emph{32},
  298--312\relax
\mciteBstWouldAddEndPuncttrue
\mciteSetBstMidEndSepPunct{\mcitedefaultmidpunct}
{\mcitedefaultendpunct}{\mcitedefaultseppunct}\relax
\EndOfBibitem
\bibitem[Andersson \latin{et~al.}(2012)Andersson, {\AA}kesson, and
  Diehl]{andersson2012}
Andersson,~J.; {\AA}kesson,~J.; Diehl,~M. {CasADi} -- {A} symbolic package for
  automatic differentiation and optimal control. Recent Advances in Algorithmic
  Differentiation. Berlin, 2012; pp 297--307\relax
\mciteBstWouldAddEndPuncttrue
\mciteSetBstMidEndSepPunct{\mcitedefaultmidpunct}
{\mcitedefaultendpunct}{\mcitedefaultseppunct}\relax
\EndOfBibitem
\bibitem[Nagy and Braatz(2003)Nagy, and Braatz]{nag03}
Nagy,~Z.~K.; Braatz,~R.~D. Robust nonlinear model predictive control of batch
  processes. \emph{AIChE Journal} \textbf{2003}, \emph{49}, 1776--1786\relax
\mciteBstWouldAddEndPuncttrue
\mciteSetBstMidEndSepPunct{\mcitedefaultmidpunct}
{\mcitedefaultendpunct}{\mcitedefaultseppunct}\relax
\EndOfBibitem
\bibitem[Srinivasan \latin{et~al.}(2003)Srinivasan, Bonvin, Visser, and
  Palanki]{sri03b}
Srinivasan,~B.; Bonvin,~D.; Visser,~E.; Palanki,~S. Dynamic optimization of
  batch processes: II. Role of measurements in handling uncertainty.
  \emph{Computers \& Chemical Engineering} \textbf{2003}, \emph{27}, 27 --
  44\relax
\mciteBstWouldAddEndPuncttrue
\mciteSetBstMidEndSepPunct{\mcitedefaultmidpunct}
{\mcitedefaultendpunct}{\mcitedefaultseppunct}\relax
\EndOfBibitem
\bibitem[Adetola \latin{et~al.}(2009)Adetola, DeHaan, and Guay]{ade09}
Adetola,~V.; DeHaan,~D.; Guay,~M. Adaptive model predictive control for
  constrained nonlinear systems. \emph{Systems \& Control Letters}
  \textbf{2009}, \emph{58}, 320 -- 326\relax
\mciteBstWouldAddEndPuncttrue
\mciteSetBstMidEndSepPunct{\mcitedefaultmidpunct}
{\mcitedefaultendpunct}{\mcitedefaultseppunct}\relax
\EndOfBibitem
\bibitem[Stuber and Barton(2011)Stuber, and Barton]{stu11}
Stuber,~M.~D.; Barton,~P.~I. Robust simulation and design using semi-infinite
  programs with implicit functions. \emph{International Journal of Reliability
  and Safety} \textbf{2011}, \emph{5}, 378--397\relax
\mciteBstWouldAddEndPuncttrue
\mciteSetBstMidEndSepPunct{\mcitedefaultmidpunct}
{\mcitedefaultendpunct}{\mcitedefaultseppunct}\relax
\EndOfBibitem
\bibitem[Francois and Bonvin(2013)Francois, and Bonvin]{fra13}
Francois,~G.; Bonvin,~D. In \emph{Control and Optimisation of Process Systems};
  Pushpavanam,~S., Ed.; Advances in Chemical Engineering Supplement C; Academic
  Press, 2013; Vol.~43; pp 1 -- 50\relax
\mciteBstWouldAddEndPuncttrue
\mciteSetBstMidEndSepPunct{\mcitedefaultmidpunct}
{\mcitedefaultendpunct}{\mcitedefaultseppunct}\relax
\EndOfBibitem
\bibitem[Lucia \latin{et~al.}(2013)Lucia, Finkler, and Engell]{luc13}
Lucia,~S.; Finkler,~T.; Engell,~S. Multi-stage nonlinear model predictive
  control applied to a semi-batch polymerization reactor under uncertainty.
  \emph{J Process Contr} \textbf{2013}, \emph{23}, 1306 -- 1319\relax
\mciteBstWouldAddEndPuncttrue
\mciteSetBstMidEndSepPunct{\mcitedefaultmidpunct}
{\mcitedefaultendpunct}{\mcitedefaultseppunct}\relax
\EndOfBibitem
\bibitem[Mart\'i \latin{et~al.}(2015)Mart\'i, Lucia, Sarabia, Paulen, Engell,
  and de~Prada]{mar15}
Mart\'i,~R.; Lucia,~S.; Sarabia,~D.; Paulen,~R.; Engell,~S.; de~Prada,~C.
  Improving scenario decomposition algorithms for robust nonlinear model
  predictive control. \emph{Computers \& Chemical Engineering} \textbf{2015},
  \emph{79}, 30--45\relax
\mciteBstWouldAddEndPuncttrue
\mciteSetBstMidEndSepPunct{\mcitedefaultmidpunct}
{\mcitedefaultendpunct}{\mcitedefaultseppunct}\relax
\EndOfBibitem
\bibitem[Jang \latin{et~al.}(2016)Jang, Lee, and Biegler]{jan16}
Jang,~H.; Lee,~J.~H.; Biegler,~L.~T. A robust NMPC scheme for semi-batch
  polymerization reactors. \emph{IFAC-PapersOnLine} \textbf{2016}, \emph{49},
  37 -- 42, 11th IFAC Symposium on Dynamics and Control of Process Systems
  Including Biosystems DYCOPS-CAB 2016\relax
\mciteBstWouldAddEndPuncttrue
\mciteSetBstMidEndSepPunct{\mcitedefaultmidpunct}
{\mcitedefaultendpunct}{\mcitedefaultseppunct}\relax
\EndOfBibitem
\bibitem[Houska \latin{et~al.}(2017)Houska, Li, and Chachuat]{hou17}
Houska,~B.; Li,~J.~C.; Chachuat,~B. Towards rigorous robust optimal control via
  generalized high-order moment expansion. \emph{Optimal Control Applications
  and Methods} \textbf{2017}, \emph{39}, 489--502\relax
\mciteBstWouldAddEndPuncttrue
\mciteSetBstMidEndSepPunct{\mcitedefaultmidpunct}
{\mcitedefaultendpunct}{\mcitedefaultseppunct}\relax
\EndOfBibitem
\bibitem[Hangos \latin{et~al.}(2006)Hangos, Bokor, and Szederk{\'e}nyi]{han06}
Hangos,~K.~M.; Bokor,~J.; Szederk{\'e}nyi,~G. \emph{Analysis and control of
  nonlinear process systems}; Springer Science \& Business Media, 2006\relax
\mciteBstWouldAddEndPuncttrue
\mciteSetBstMidEndSepPunct{\mcitedefaultmidpunct}
{\mcitedefaultendpunct}{\mcitedefaultseppunct}\relax
\EndOfBibitem
\bibitem[Sontag(1998)]{son98}
Sontag,~E.~D. \emph{Mathematical Control Theory: Deterministic Finite
  Dimensional Systems (2nd Ed.)}; Springer-Verlag: Berlin, Heidelberg,
  1998\relax
\mciteBstWouldAddEndPuncttrue
\mciteSetBstMidEndSepPunct{\mcitedefaultmidpunct}
{\mcitedefaultendpunct}{\mcitedefaultseppunct}\relax
\EndOfBibitem
\bibitem[Amrhein \latin{et~al.}(2010)Amrhein, Bhatt, Srinivasan, and
  Bonvin]{amr10}
Amrhein,~M.; Bhatt,~N.; Srinivasan,~B.; Bonvin,~D. Extents of Reaction and Flow
  for Homogeneous Reaction Systems with Inlet and Outlet Streams. \emph{AIChE
  Journal} \textbf{2010}, \emph{56}\relax
\mciteBstWouldAddEndPuncttrue
\mciteSetBstMidEndSepPunct{\mcitedefaultmidpunct}
{\mcitedefaultendpunct}{\mcitedefaultseppunct}\relax
\EndOfBibitem
\bibitem[Liou and Hsiue(1995)Liou, and Hsiue]{lio95}
Liou,~C.; Hsiue,~T. Exact linearization and control of a continuous stirred
  tank reactor. \emph{Journal of the Chinese Institute of Engineers}
  \textbf{1995}, \emph{18}, 825--833\relax
\mciteBstWouldAddEndPuncttrue
\mciteSetBstMidEndSepPunct{\mcitedefaultmidpunct}
{\mcitedefaultendpunct}{\mcitedefaultseppunct}\relax
\EndOfBibitem
\bibitem[Feldbaum(1995)]{fel60}
Feldbaum,~A. Dual control theory.~{I}. \emph{Avtomatika i telemekhanika}
  \textbf{1995}, \emph{21}, 1240--1249\relax
\mciteBstWouldAddEndPuncttrue
\mciteSetBstMidEndSepPunct{\mcitedefaultmidpunct}
{\mcitedefaultendpunct}{\mcitedefaultseppunct}\relax
\EndOfBibitem
\bibitem[{Filatov} and {Unbehauen}(2000){Filatov}, and {Unbehauen}]{fil00}
{Filatov},~N.~M.; {Unbehauen},~H. Survey of adaptive dual control methods.
  \emph{IEE Proceedings - Control Theory and Applications} \textbf{2000},
  \emph{147}, 118--128\relax
\mciteBstWouldAddEndPuncttrue
\mciteSetBstMidEndSepPunct{\mcitedefaultmidpunct}
{\mcitedefaultendpunct}{\mcitedefaultseppunct}\relax
\EndOfBibitem
\bibitem[{Unbehauen}(2000)]{unb00}
{Unbehauen},~H. Adaptive dual control systems: a survey. Proceedings of the
  IEEE 2000 Adaptive Systems for Signal Processing, Communications, and Control
  Symposium (Cat. No.00EX373). 2000; pp 171--180\relax
\mciteBstWouldAddEndPuncttrue
\mciteSetBstMidEndSepPunct{\mcitedefaultmidpunct}
{\mcitedefaultendpunct}{\mcitedefaultseppunct}\relax
\EndOfBibitem
\bibitem[{Milito} \latin{et~al.}(1982){Milito}, {Padilla}, {Padilla}, and
  {Cadorin}]{mil82}
{Milito},~R.; {Padilla},~C.; {Padilla},~R.; {Cadorin},~D. An innovations
  approach to dual control. \emph{IEEE Transactions on Automatic Control}
  \textbf{1982}, \emph{27}, 132--137\relax
\mciteBstWouldAddEndPuncttrue
\mciteSetBstMidEndSepPunct{\mcitedefaultmidpunct}
{\mcitedefaultendpunct}{\mcitedefaultseppunct}\relax
\EndOfBibitem
\bibitem[Hanssen and Foss(2015)Hanssen, and Foss]{han15}
Hanssen,~K.~G.; Foss,~B. Scenario Based Implicit Dual Model Predictive Control.
  \emph{IFAC-PapersOnLine} \textbf{2015}, \emph{48}, 416 -- 421, 5th IFAC
  Conference on Nonlinear Model Predictive Control NMPC 2015\relax
\mciteBstWouldAddEndPuncttrue
\mciteSetBstMidEndSepPunct{\mcitedefaultmidpunct}
{\mcitedefaultendpunct}{\mcitedefaultseppunct}\relax
\EndOfBibitem
\bibitem[Heirung \latin{et~al.}(2015)Heirung, Foss, and Ydstie]{hei15}
Heirung,~T. A.~N.; Foss,~B.; Ydstie,~B.~E. MPC-based dual control with online
  experiment design. \emph{Journal of Process Control} \textbf{2015},
  \emph{32}, 64 -- 76\relax
\mciteBstWouldAddEndPuncttrue
\mciteSetBstMidEndSepPunct{\mcitedefaultmidpunct}
{\mcitedefaultendpunct}{\mcitedefaultseppunct}\relax
\EndOfBibitem
\bibitem[La \latin{et~al.}(2017)La, Potschka, Schl{\"o}der, and Bock]{la17}
La,~H.~C.; Potschka,~A.; Schl{\"o}der,~J.~P.; Bock,~H.~G. Dual Control and
  Online Optimal Experimental Design. \emph{SIAM Journal on Scientific
  Computing} \textbf{2017}, \emph{39}, B640--B657\relax
\mciteBstWouldAddEndPuncttrue
\mciteSetBstMidEndSepPunct{\mcitedefaultmidpunct}
{\mcitedefaultendpunct}{\mcitedefaultseppunct}\relax
\EndOfBibitem
\bibitem[Lorenzen \latin{et~al.}(2019)Lorenzen, Cannon, and Allgöwer]{lor19}
Lorenzen,~M.; Cannon,~M.; Allgöwer,~F. Robust MPC with recursive model update.
  \emph{Automatica} \textbf{2019}, \emph{103}, 461--471\relax
\mciteBstWouldAddEndPuncttrue
\mciteSetBstMidEndSepPunct{\mcitedefaultmidpunct}
{\mcitedefaultendpunct}{\mcitedefaultseppunct}\relax
\EndOfBibitem
\bibitem[Lee and Lee(2009)Lee, and Lee]{lee09}
Lee,~J.~M.; Lee,~J.~H. An approximate dynamic programming based approach to
  dual adaptive control. \emph{Journal of Process Control} \textbf{2009},
  \emph{19}, 859 -- 864\relax
\mciteBstWouldAddEndPuncttrue
\mciteSetBstMidEndSepPunct{\mcitedefaultmidpunct}
{\mcitedefaultendpunct}{\mcitedefaultseppunct}\relax
\EndOfBibitem
\bibitem[Tanaskovic \latin{et~al.}(2014)Tanaskovic, Fagiano, Smith, and
  Morari]{tan14}
Tanaskovic,~M.; Fagiano,~L.; Smith,~R.; Morari,~M. Adaptive receding horizon
  control for constrained MIMO systems. \emph{Automatica} \textbf{2014},
  \emph{50}, 3019 -- 3029\relax
\mciteBstWouldAddEndPuncttrue
\mciteSetBstMidEndSepPunct{\mcitedefaultmidpunct}
{\mcitedefaultendpunct}{\mcitedefaultseppunct}\relax
\EndOfBibitem
\bibitem[Tanaskovic \latin{et~al.}(2019)Tanaskovic, Fagiano, and
  Gligorovski]{tan19}
Tanaskovic,~M.; Fagiano,~L.; Gligorovski,~V. Adaptive model predictive control
  for linear time varying MIMO systems. \emph{Automatica} \textbf{2019},
  \emph{105}, 237 -- 245\relax
\mciteBstWouldAddEndPuncttrue
\mciteSetBstMidEndSepPunct{\mcitedefaultmidpunct}
{\mcitedefaultendpunct}{\mcitedefaultseppunct}\relax
\EndOfBibitem
\bibitem[Heirung \latin{et~al.}(2017)Heirung, Ydstie, and Foss]{hei17}
Heirung,~T. A.~N.; Ydstie,~B.~E.; Foss,~B. Dual adaptive model predictive
  control. \emph{Automatica} \textbf{2017}, \emph{80}, 340 -- 348\relax
\mciteBstWouldAddEndPuncttrue
\mciteSetBstMidEndSepPunct{\mcitedefaultmidpunct}
{\mcitedefaultendpunct}{\mcitedefaultseppunct}\relax
\EndOfBibitem
\bibitem[Feng and Houska(2018)Feng, and Houska]{fen18}
Feng,~X.; Houska,~B. Real-time algorithm for self-reflective model predictive
  control. \emph{Journal of Process Control} \textbf{2018}, \emph{65},
  68--77\relax
\mciteBstWouldAddEndPuncttrue
\mciteSetBstMidEndSepPunct{\mcitedefaultmidpunct}
{\mcitedefaultendpunct}{\mcitedefaultseppunct}\relax
\EndOfBibitem
\bibitem[Thangavel \latin{et~al.}(2015)Thangavel, Lucia, Paulen, and
  Engell]{tha15}
Thangavel,~S.; Lucia,~S.; Paulen,~R.; Engell,~S. Towards dual robust nonlinear
  model predictive control: A multi-stage approach. Proc. Amer Contr Conf.
  2015; pp 428--433\relax
\mciteBstWouldAddEndPuncttrue
\mciteSetBstMidEndSepPunct{\mcitedefaultmidpunct}
{\mcitedefaultendpunct}{\mcitedefaultseppunct}\relax
\EndOfBibitem
\bibitem[Thangavel \latin{et~al.}(2018)Thangavel, Lucia, Paulen, and
  Engell]{tha18}
Thangavel,~S.; Lucia,~S.; Paulen,~R.; Engell,~S. Dual robust nonlinear model
  predictive control: A multi-stage approach. \emph{Journal of Process Control}
  \textbf{2018}, \emph{72}, 39 -- 51\relax
\mciteBstWouldAddEndPuncttrue
\mciteSetBstMidEndSepPunct{\mcitedefaultmidpunct}
{\mcitedefaultendpunct}{\mcitedefaultseppunct}\relax
\EndOfBibitem
\bibitem[Pontryagin \latin{et~al.}(1962)Pontryagin, Boltyanskii, Gamkrelidze,
  and Mishchenko]{pon62}
Pontryagin,~L.~S.; Boltyanskii,~V.~G.; Gamkrelidze,~R.~V.; Mishchenko,~E.~F.
  \emph{{The Mathematical Theory of Optimal Processes}}; John Wiley \& Sons,
  Inc.: New York, 1962\relax
\mciteBstWouldAddEndPuncttrue
\mciteSetBstMidEndSepPunct{\mcitedefaultmidpunct}
{\mcitedefaultendpunct}{\mcitedefaultseppunct}\relax
\EndOfBibitem
\bibitem[Schweppe(1968)]{schw68}
Schweppe,~F. Recursive state estimation: Unknown but bounded errors and system
  inputs. \emph{IEEE Transactions on Automatic Control} \textbf{1968},
  \emph{13}, 22--28\relax
\mciteBstWouldAddEndPuncttrue
\mciteSetBstMidEndSepPunct{\mcitedefaultmidpunct}
{\mcitedefaultendpunct}{\mcitedefaultseppunct}\relax
\EndOfBibitem
\bibitem[Fogel and Huang(1982)Fogel, and Huang]{fog82}
Fogel,~E.; Huang,~Y. On the value of information in system identification --
  Bounded noise case. \emph{Automatica} \textbf{1982}, \emph{18}, 229 --
  238\relax
\mciteBstWouldAddEndPuncttrue
\mciteSetBstMidEndSepPunct{\mcitedefaultmidpunct}
{\mcitedefaultendpunct}{\mcitedefaultseppunct}\relax
\EndOfBibitem
\bibitem[Paulen \latin{et~al.}(2015)Paulen, Jelemensk\'y, Kov\'acs, and
  Fikar]{pau15jpc}
Paulen,~R.; Jelemensk\'y,~M.; Kov\'acs,~Z.; Fikar,~M. Economically optimal
  batch diafiltration via analytical multi-objective optimal control.
  \emph{Journal of Process Control} \textbf{2015}, \emph{28}, 73 -- 82\relax
\mciteBstWouldAddEndPuncttrue
\mciteSetBstMidEndSepPunct{\mcitedefaultmidpunct}
{\mcitedefaultendpunct}{\mcitedefaultseppunct}\relax
\EndOfBibitem
\bibitem[Aydin \latin{et~al.}(2018)Aydin, Bonvin, and Sundmacher]{erd18}
Aydin,~E.; Bonvin,~D.; Sundmacher,~K. Toward Fast Dynamic Optimization: An
  Indirect Algorithm That Uses Parsimonious Input Parameterization.
  \emph{Industrial \& Engineering Chemistry Research} \textbf{2018}, \emph{57},
  10038--10048\relax
\mciteBstWouldAddEndPuncttrue
\mciteSetBstMidEndSepPunct{\mcitedefaultmidpunct}
{\mcitedefaultendpunct}{\mcitedefaultseppunct}\relax
\EndOfBibitem
\bibitem[Rodrigues and Bonvin(0)Rodrigues, and Bonvin]{rod19}
Rodrigues,~D.; Bonvin,~D. Dynamic Optimization of Reaction Systems via Exact
  Parsimonious Input Parameterization. \emph{Industrial \& Engineering
  Chemistry Research} \textbf{0}, \emph{0}, null\relax
\mciteBstWouldAddEndPuncttrue
\mciteSetBstMidEndSepPunct{\mcitedefaultmidpunct}
{\mcitedefaultendpunct}{\mcitedefaultseppunct}\relax
\EndOfBibitem
\bibitem[Villanueva \latin{et~al.}(2015)Villanueva, Houska, and
  Chachuat]{vil15}
Villanueva,~M.~E.; Houska,~B.; Chachuat,~B. Unified framework for the
  propagation of continuous-time enclosures for parametric nonlinear ODEs.
  \emph{Journal of Global Optimization} \textbf{2015}, \emph{62},
  575--613\relax
\mciteBstWouldAddEndPuncttrue
\mciteSetBstMidEndSepPunct{\mcitedefaultmidpunct}
{\mcitedefaultendpunct}{\mcitedefaultseppunct}\relax
\EndOfBibitem
\bibitem[Lucia and Paulen(2014)Lucia, and Paulen]{lucia2014b}
Lucia,~S.; Paulen,~R. Robust Nonlinear Model Predictive Control with Reduction
  of Uncertainty via Robust Optimal Experiment Design. \emph{IFAC-PapersOnLine}
  \textbf{2014}, \emph{47}, 1904 -- 1909\relax
\mciteBstWouldAddEndPuncttrue
\mciteSetBstMidEndSepPunct{\mcitedefaultmidpunct}
{\mcitedefaultendpunct}{\mcitedefaultseppunct}\relax
\EndOfBibitem
\bibitem[Lee and Yu(1997)Lee, and Yu]{lee97}
Lee,~J.; Yu,~Z. Worst-case formulations of model predictive control for systems
  with bounded parameters. \emph{Automatica} \textbf{1997}, \emph{33}, 763 --
  781\relax
\mciteBstWouldAddEndPuncttrue
\mciteSetBstMidEndSepPunct{\mcitedefaultmidpunct}
{\mcitedefaultendpunct}{\mcitedefaultseppunct}\relax
\EndOfBibitem
\bibitem[Yousfi \latin{et~al.}(2017)Yousfi, Raïssi, Amairi, and Aoun]{you17}
Yousfi,~B.; Raïssi,~T.; Amairi,~M.; Aoun,~M. Set-membership methodology for
  model-based prognosis. \emph{ISA Transactions} \textbf{2017}, \emph{66}, 216
  -- 225\relax
\mciteBstWouldAddEndPuncttrue
\mciteSetBstMidEndSepPunct{\mcitedefaultmidpunct}
{\mcitedefaultendpunct}{\mcitedefaultseppunct}\relax
\EndOfBibitem
\bibitem[de~Prada \latin{et~al.}(2011)de~Prada, Rodriguez, and Sarabia]{pra11}
de~Prada,~C.; Rodriguez,~M.; Sarabia,~D. On-Line Scheduling and Control of a
  Mixed Continuous-Batch Plant. \emph{Industrial \& Engineering Chemistry
  Research} \textbf{2011}, \emph{50}, 5041--5049\relax
\mciteBstWouldAddEndPuncttrue
\mciteSetBstMidEndSepPunct{\mcitedefaultmidpunct}
{\mcitedefaultendpunct}{\mcitedefaultseppunct}\relax
\EndOfBibitem
\bibitem[Lee and Lee(2005)Lee, and Lee]{LEE20051281}
Lee,~J.~M.; Lee,~J.~H. Approximate dynamic programming-based approaches for
  input–output data-driven control of nonlinear processes. \emph{Automatica}
  \textbf{2005}, \emph{41}, 1281--1288\relax
\mciteBstWouldAddEndPuncttrue
\mciteSetBstMidEndSepPunct{\mcitedefaultmidpunct}
{\mcitedefaultendpunct}{\mcitedefaultseppunct}\relax
\EndOfBibitem
\bibitem[Paulen \latin{et~al.}(2012)Paulen, Fikar, Foley, Kov{\'a}cs, and
  Czermak]{pau12jms}
Paulen,~R.; Fikar,~M.; Foley,~G.; Kov{\'a}cs,~Z.; Czermak,~P. {Opti\-mal
  feeding strategy of diafiltration buffer in batch membrane processes}.
  \emph{Journal of Membrane Science} \textbf{2012}, \emph{411-412},
  160--172\relax
\mciteBstWouldAddEndPuncttrue
\mciteSetBstMidEndSepPunct{\mcitedefaultmidpunct}
{\mcitedefaultendpunct}{\mcitedefaultseppunct}\relax
\EndOfBibitem
\bibitem[S{\"o}derstr{\"o}m(2007)]{sod07}
S{\"o}derstr{\"o}m,~T. Errors-in-variables methods in system identification.
  \emph{Automatica} \textbf{2007}, \emph{43}, 939 -- 958\relax
\mciteBstWouldAddEndPuncttrue
\mciteSetBstMidEndSepPunct{\mcitedefaultmidpunct}
{\mcitedefaultendpunct}{\mcitedefaultseppunct}\relax
\EndOfBibitem
\bibitem[Gottu~Mukkula and Paulen(2017)Gottu~Mukkula, and Paulen]{mupa_cace17}
Gottu~Mukkula,~A.~R.; Paulen,~R. Model-based design of optimal experiments for
  nonlinear systems in the context of guaranteed parameter estimation.
  \emph{Computers \& Chemical Engineering} \textbf{2017}, \emph{99}, 198 --
  213\relax
\mciteBstWouldAddEndPuncttrue
\mciteSetBstMidEndSepPunct{\mcitedefaultmidpunct}
{\mcitedefaultendpunct}{\mcitedefaultseppunct}\relax
\EndOfBibitem
\bibitem[Srinivasan \latin{et~al.}(2003)Srinivasan, Palanki, and
  Bonvin]{sri03a}
Srinivasan,~B.; Palanki,~S.; Bonvin,~D. {{D}ynamic optimization of batch
  processes: {I}. {C}haracterization of the nominal solution}. \emph{Computers
  \& Chemical Engineering} \textbf{2003}, \emph{27}, 1--26\relax
\mciteBstWouldAddEndPuncttrue
\mciteSetBstMidEndSepPunct{\mcitedefaultmidpunct}
{\mcitedefaultendpunct}{\mcitedefaultseppunct}\relax
\EndOfBibitem
\bibitem[J{\"o}nsson and Tr{\"a}g{\aa}rdh(1990)J{\"o}nsson, and
  Tr{\"a}g{\aa}rdh]{jon90}
J{\"o}nsson,~A.-S.; Tr{\"a}g{\aa}rdh,~G. {Ultrafiltration applications}.
  \emph{Desalination} \textbf{1990}, \emph{77}, 135 -- 179, {P}roceedings of
  the Symposium on Membrane Technology\relax
\mciteBstWouldAddEndPuncttrue
\mciteSetBstMidEndSepPunct{\mcitedefaultmidpunct}
{\mcitedefaultendpunct}{\mcitedefaultseppunct}\relax
\EndOfBibitem
\end{mcitethebibliography}
\end{document}